\documentclass[aps,pra,twocolumn,floatfix]{revtex4}
\usepackage{amssymb}
\usepackage{amsmath}
\usepackage{bm}
\usepackage{epsfig}

\def\be{\begin{equation}}
\def\ee{\end{equation}}
\def\bea{\begin{eqnarray}}
\def\eea{\end{eqnarray}}

\begin{document}

\title{Condensation of N bosons IV: A simplified Bogoliubov master equation
analysis of fluctuations in an interacting Bose gas}
\author{Marlan O. Scully and Anatoly A. Svidzinsky}
\affiliation{{\small Applied Physics and Materials Science Group, Eng. Quad., Princeton
Univ., Princeton 08544}\\
Institute for Quantum Studies and Dept. of Physics, Texas A\&M Univ., Texas
77843}
\date{\today }

\begin{abstract}
A nonequilibrium master equation analysis for $N$ interacting bosons, with
Bogoliubov quasiparticles as the reservoir is presented. The analysis is
based on a simplified Hamiltonian. The steady state solution yields the
equilibrium density matrix. The results are in good agreement with and
extend our previous rigorous canonical ensemble equilibrium statistical
treatment leading to a quantum theory of the atom laser.
\end{abstract}

\maketitle

\section{Introduction}

\begin{figure}[h]
\bigskip 
\centerline{\epsfxsize=0.55\textwidth\epsfysize=0.45\textwidth
\epsfbox{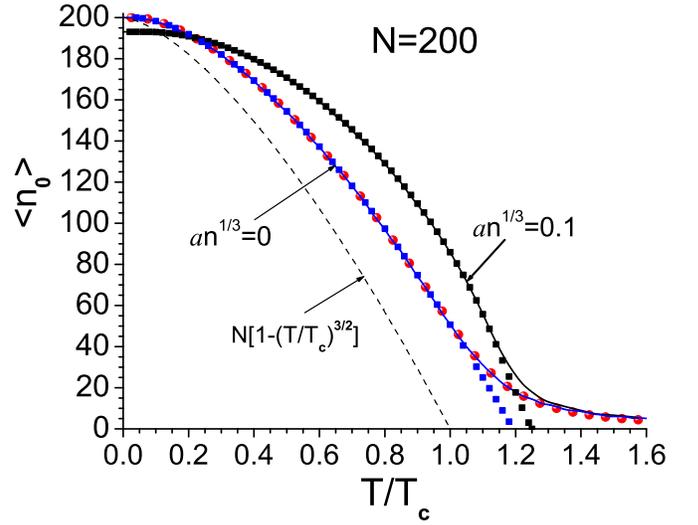}}
\caption{Solid lines are the present work results showing the mean number of
condensate particles as a function of temperature for N = 200 particles in a
box calculated via the solution of the condensate master equation for an
interacting (Bogoliubov) gas with $an^{1/3}$ = 0.1 and an ideal (Bose) gas $%
an^{1/3}$ = 0. Dots are the exact numerical results obtained in the
canonical ensemble for the ideal Bose gas \protect\cite{la}. Squares are the
results of CNB3 (please note that repulsive interaction increases $T_c$).
Dashed line is a plot of $N[1-(T/T_{c})^{3/2}]$ which is valid for the ideal
gas in the thermodynamic limit. The temperature is normalized by the ideal
gas thermodynamic critical temperature in the box $T_{c}=2\protect\pi %
\hbar^2 n^{2/3}/k_{B}M\protect\zeta (3/2)^{2/3}$, where $M$ is the particle
mass. }
\label{n0pqe}
\end{figure}

We here analyze the condensate fluctuations of $N$ interacting bosons via a
master equation analysis based on a simple truncated Hamiltonian. The
strategy is to regard excited states of the Bose gas, as well as collective
excitations, as reservoir variables which we ultimately trace over, in order
to obtain a master equation for the condensate. We find that the steady
state solution of the resulting kinetic equations provides a good
description of Bose-Einstein condensate (BEC) \cite{Meys01}. Fig. \ref{n0pqe}
shows the condensate fraction as a function of temperature and is one of our
main results. This is in accord with the wit and wisdom of Wigner:

\begin{quotation}
\noindent\emph{{With classical thermodynamics, one can calculate almost
everything crudely; with kinetic theory, one can calculate fewer things, but
more accurately; and with statistical mechanics, one can calculate almost
nothing exactly.} }
\end{quotation}

In particular, we discuss several subtle issues concerning the statistics of
condensate atoms in the region of $T\simeq T_{c} $. This is analogous to
studying the photon statistics of the laser in the passage from below to
above threshold. In fact, BEC is often referred to as an atom laser, and the
question of the atom statistics of the mesoscopic BEC (containing a few 10's
or 100's of atoms), is the subject of this paper. The answer is given by the
diagonal elements of the BEC density matrix, $\rho _{n_{0},n_{0}}$, where $%
n_{0}$ is the number of atoms in the BEC ground state. In the present case
of a gas of $N$ weakly interacting bosons, in one limit we find

\begin{equation}
\rho _{n_{0},n_{0}}=\frac{\mathcal{H}_{I}^{N-n_{0}}}{(N-n_{0})!}{\mathcal{Z}}%
_{N}^{-1},\qquad \mathcal{Z}_{N}=e^{\mathcal{H}_{I}}\Gamma (N+1,\mathcal{H}%
_{I})/N!  \label{rho}
\end{equation}%
\newline
where $\mathcal{Z}_{N}$ is a normalization (partition function) factor
expressed in terms of the incomplete gamma function $\Gamma (N+1,\mathcal{H}%
_{I})$; and $\mathcal{H}_{I}$ is a simple heating coefficient governing the
rate of removal of atoms from the ground state due to scattering by
Bogoliubov quasiparticles, which is given by 
\begin{equation}
\mathcal{H}_{I}={\sum_{\mathbf{k\neq 0}}}\left( \frac{u_{\mathbf{k}}^{2}+v_{%
\mathbf{k}}^{2}}{e^{\beta \varepsilon _{\mathbf{k}}}-1}+v_{\mathbf{k}%
}^{2}\right) \text{.}
\end{equation}

In the above the usual Bogoliubov amplitudes are 
\begin{eqnarray}
u_{\mathbf{k}} &=&\frac{1}{\sqrt{1-A_{\mathbf{k}}^{2}}},\;v_{\mathbf{k}}=%
\frac{A_{\mathbf{k}}}{\sqrt{1-A_{\mathbf{k}}^{2}}},  \notag \\
A_{\mathbf{k}} &=&\frac{V}{{\bar{n}_{0}}U_{\mathbf{k}}}\left( \epsilon _{%
\mathbf{k}}-\frac{\hbar ^{2}k^{2}}{2M}-\frac{{\bar{n}_{0}}U_{\mathbf{k}}}{V}%
\right) ,  \label{Ak}
\end{eqnarray}
where $M$ is the atomic mass, $V$ is the condensate volume and $U_{\mathbf{k}%
}$ is the atom-atom scattering energy; the quasiparticle energy $\epsilon _{%
\mathbf{k}}$ is given by 
\begin{equation}
\epsilon _{\mathbf{k}}=\sqrt{\left( \frac{\hbar ^{2}k^{2}}{2M}+\frac{{\bar{n}%
_{0}}U_{\mathbf{k}}}{V}\right) ^{2}-\left( \frac{{\bar{n}_{0}}U_{\mathbf{k}}%
}{V}\right) ^{2}}.  \label{ek}
\end{equation}

The simple expression, Eq.~(\ref{rho}), is valid in the limit in which, for
the case of an ideal gas in a box, $\mathcal{H}_{I}=N\left( T/T_{c}\right)
^{3/2}$ where $T$ is the temperature of the gas and $T_{c}$ is the usual
critical temperature. This was the main result of the first paper on the
Condensation of N Bosons (CNB1) in this series \cite{MOS99}, in which we
focused on the statistics of $N$ noninteracting (ideal) bosons. In the
second paper \cite{KSZZ} (CNB2) the approximations were improved, as
discussed in section III below. In CNB1,2 we solved for the condensate
statistics of an ideal Bose gas and found, surprisingly, that the
fluctuations are not Gaussian even in the thermodynamic limit. The treatment
of CNB1,2 is patterned after the quantum theory of the laser, i.e., is based
on a master equation for the condensate density matrix. As such, it has the
useful (and intriguing) feature of working equally well at all temperatures
from $T=0$ to $T_{c}$ and above. Atom-atom interactions are, however,
ignored in those papers.

\begin{figure}[h]
\bigskip 
\centerline{\epsfxsize=0.55\textwidth\epsfysize=0.45\textwidth
\epsfbox{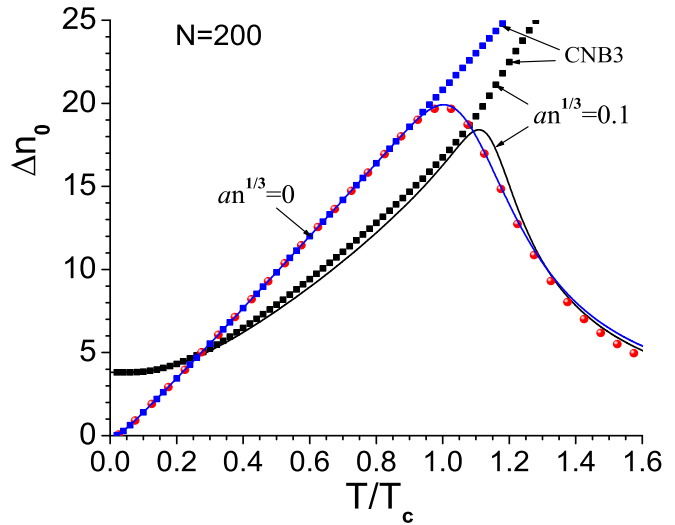}}
\caption{Variance $\Delta n_{0}$ of the condensate particle number as a
function of temperature for N = 200 particles in a box calculated via the
solution of the condensate master equation for $an^{1/3}$ = 0 and 0.1 (solid
lines). Dots are obtained by exact numerical calculations for the ideal Bose
gas in the canonical ensemble \protect\cite{la}. Squares are the result of
CNB3. For $an^{1/3}$ = 0.1 the solid line is obtained from the master
equation by a shift of $\mathcal{H}_I$ to match the CNB3 value at T = 0 (see
Table 1 column three).}
\label{v2pqe}
\end{figure}

Paper CNB3 \cite{KKS} includes interactions. The analysis of CNB3 is based
on the Girardeau-Arnowitt adaptation of Bogoliubov's approach to BEC. In
CNB3, we give a \textquotedblleft rigorous\textquotedblright\ statistical
mechanical treatment of the fluctuations, valid provided the average number
of condensate particles is much larger than its variance; that is valid for
temperatures not too near $T_{c}$. This condition implies that the excited
atom numbers $n_{k}$ fluctuate independently, $<n_{k}n_{m}>=\bar{n}_{k}\bar{n%
}_{m}$, $k\neq m$, by exchanging particles with the condensate reservoir.
For $N=200$ CNB3 is valid from $T=0$ to $T\approx 0.7T_{c}$. However, near
and above $T_{c}$ the correlations become substantial which yields failure
of such a treatment. In a broad temperature range, interactions increase the
condensate number $\left\langle n_{0}\right\rangle $, and the fluctuations
are found to be one half those of the ideal Bose gas. This remarkable fact
is due to the $\left( \vec{k}-\vec{k}\right) $ pairing of atoms.

In the present paper, we utilize a master equation analysis for the
interacting Bose gas based on the simple truncated Hamiltonian given by Eq. (%
\ref{V_quasi}). In particular we derive the average number of particles in
the lowest energy level, $\langle n_{0}\rangle $ and the fluctuation about
the average $\Delta n_{0}^{2}=\langle (n_{0}-\langle n_{0}\rangle )^{2}$.
These are plotted in Figs. \ref{n0pqe} and \ref{v2pqe} and constitute our
main results.

We note that the present analysis yields an expression for $\left\langle
n_{0}\right\rangle $, in good agreement with the rigorous results of CNB3,
in the temperature region where that analysis \ is valid; and extends those
results to higher temperature up to $T_{c}$ and beyond, as per Fig. \ref%
{n0pqe}.

We emphasize that the problem of the interacting Bose gas near $T_{c}$ is a
difficult one. It is therefore worthwhile to investigate simple models
designed to capture the essential physics. To that end, the next section is
based on a simple \textquotedblleft toy\textquotedblright\ model which works
surprisingly well. Encouraged by the results of that analysis we then
present a physically motivated model yielding a similar Hamiltonian. The
analysis based on a more general treatment will be published elsewhere.

In the next section we derive the relevant master equation which we solve in
section III to obtain the condensate statistics. In section IV we discuss
the results and make contact with previous work.

\section{Master equation for weakly interacting Bose gas}

Having set the stage, we proceed to sketch a simple approach extending our
master equation analysis to the case of the interacting Bose gas. The
present treatment is based on simple extension of CNB1,2 via a ``toy'' model
and then a more realistic (but still heuristic) formulation. However the
results are in good agreement with ``rigorous''\ results of CNB3 for
temperatures where that theory is valid and provide answers in the vicinity
of $T_{c}$ where it does breakdown.

In CNB1,2 the basic model interaction Hamiltonian describing the statistical
dynamics of the condensate was taken to be 
\begin{equation}
V=\sum_{\mathbf{k}}g_{\mathbf{k}}\,{\hat{a}}_{0}^{\dagger }\,{\hat{b}}_{%
\mathbf{k}}^{\dagger }\,{\hat{a}}_{\mathbf{k}}+\mathrm{adj.}  \label{V}
\end{equation}
where ${\hat{a}}_{\mathbf{k}}$ and ${\hat{b}}_{\mathbf{k}}$ are atom and
phonon annihilation operators, likewise ${\hat{a}}_{0}$ is the condensate
annihilation operator and $g_{\mathbf{k}}$ is the corresponding coupling
strength for the collision of a ground state atom and an atom having
momentum $\mathbf{k}$ scattering into the BEC. This models the dilute gas
BEC experiments of Reppy and coworkers \cite{rep}. Moreover the results for
the condensate particle number, obtained by the master equation approach of
CNB2 and exact numerical simulation for the ideal gas in the canonical
ensemble \cite{la}, are in excellent agreement as indicated in Fig. \ref%
{n0pqe}.

\subsection{Toy Model for an Interacting Bose Gas}

As a first step toward obtaining $\bar{n}_{0}$ and $\Delta n_{0}$ for an
interacting Bose gas, we derive a master equation for $\rho _{n_{0},n_{0}}$
by simply replacing the phonon operators $b_{\mathbf{k}}$ and $b_{\mathbf{k}%
}^{\dagger }$ in Eq. (\ref{V}) by operators \^{\ss }$_{\mathbf{k}}$ and \^{%
\ss }$_{\mathbf{k}}^{\dagger }$ related to $\hat{b}_{\mathbf{k}}$ and $\hat{b%
}_{\mathbf{k}}^{\dagger }$ by the usual Bogoliubov transformation%
\begin{eqnarray}
\text{\^{\ss }}_{\mathbf{k}} &=&u_{\mathbf{k}}\hat{b}_{\mathbf{k}}+v_{%
\mathbf{k}}\hat{b}_{-\mathbf{k}}^{\dagger }\text{ \ \ }  \notag \\
\text{\^{\ss }}_{\mathbf{k}}^{\dagger } &=&u_{\mathbf{k}}\hat{b}_{\mathbf{k}%
}^{\dagger }+v_{\mathbf{k}}\hat{b}_{-\mathbf{k}},  \label{BG}
\end{eqnarray}%
where $u_{\mathbf{k}}$ and $v_{\mathbf{k}}$ are given in Eq. (\ref{Ak}), $<%
\hat{b}_{\mathbf{k}}^{\dagger }\hat{b}_{\mathbf{k}}>=1/(e^{\beta \epsilon _{%
\mathbf{k}}}-1)$, $\epsilon _{\mathbf{k}}$ is the Bogoliubov quasiparticle
energy given by Eq. (\ref{ek}).

The present master equation for the condensate is now obtained via the
phenomenological interaction Hamiltonian 
\begin{equation}
V=\sum_{\mathbf{k}}\tilde{g}_{\mathbf{k}}\,{\hat{a}}_{\mathbf{k}}\,\text{\^{%
\ss }}_{\mathbf{k}}^{\dagger }\,{\hat{a}}_{0}^{\dagger }+\mathrm{adj.}
\label{V3}
\end{equation}%
which adds (removes) atoms from the condensate with the annihilation
(creation) of an excited atom and the emission (absorption) of a
quasiparticle.

We recall that the simple ideal gas master equation for the probability $%
P_{n_{0}}$ of finding $n_{0}$ atoms in the N atom gas ground state was found
to be \cite{MOS99}%
\begin{equation*}
\frac{1}{\kappa }\dot{P}%
_{n_{0}}=-(N-n_{0})(n_{0}+1)P_{n_{0}}+(N-n_{0}+1)n_{0}P_{n_{0}-1}-
\end{equation*}
\begin{equation}
\mathcal{H}n_{0}P_{n_{0}}+\mathcal{H}(n_{0}+1)P_{n_{0}+1}\text{,}
\end{equation}%
where $\kappa $ is an uninteresting overall rate factor and the phonon
heating coefficient was given by%
\begin{equation}
\mathcal{H}={\sum_{\mathbf{k\neq 0}}}\frac{1}{e^{\beta \epsilon _{\mathbf{k}%
}}-1}\text{,}
\end{equation}%
where $\beta =1/k_{B}T$.

In the present toy model all we do is replace the phonon operators $\hat{b}_{%
\mathbf{k}}$ and $\hat{b}_{\mathbf{k}}^{\dagger }$ in Eq. (\ref{V}) by the
quasiparticle operators \^{\ss }$_{\mathbf{k}}$ and \^{\ss }$_{\mathbf{k}%
}^{\dagger }$ of Eq. (\ref{BG}), and find the new heating coefficient 
\begin{equation}
\mathcal{H}\Rightarrow {\sum_{\mathbf{k\neq 0}}}\left[ \frac{%
u_{k}^{2}+v_{k}^{2}}{e^{\beta \epsilon _{\mathbf{k}}}-1}+v_{k}^{2}\right]
\equiv \mathcal{H}_{I}\text{.}
\end{equation}

The results of this simple analysis are summarized in Figs. \ref{n0pqe} and %
\ref{v2pqe}. Encouraged by these results we next present a simple argument
which puts the phenomenological analysis of this subsection on a more
physically motivated footing.

\subsection{A Physical Argument Based on Bogoliubov-Girardeau-Arnowitt
Formalism}

We work with the particle number conserving creation and annihilation
operators of Girardeau and Arnowitt \cite{ga} 
\begin{equation}
\text{$\hat{\beta}$}_{\mathbf{k}}^{\dagger }={\hat{a}}_{\mathbf{k}}^{\dagger
}\text{$\hat{\beta}$}_{0}\text{, \ \ \ \ $\hat{\beta}$}_{\mathbf{k}}=\text{$%
\hat{\beta}$}_{0}^{\dagger }{\hat{a}}_{\mathbf{k}}\text{, \ \ \ \ $\hat{\beta%
}$}_{0}=(1+\hat{n}_{0})^{-1/2}{\hat{a}}_{0}\text{{.}}
\end{equation}%
These operators describe canonical-ensemble quasiparticles which obey the
Bose canonical commutation relations 
\begin{equation}
\lbrack \text{$\hat{\beta}$}_{\mathbf{k}},\text{$\hat{\beta}$}_{\mathbf{k}%
}^{\dagger }]=\delta _{\mathbf{k},\mathbf{k}^{\prime }},
\end{equation}%
see CNB3. The canonical-ensemble quasiparticles $\hat{\beta}_{\mathbf{k}}=%
\hat{\beta}_{0}^{\dagger }{\hat{a}}_{\mathbf{k}}$ describe transitions
between ground ($\mathbf{k=0}$) and excited ($\mathbf{k}\neq \mathbf{0}$)
states.

\begin{figure}[h]
\bigskip 
\centerline{\epsfxsize=0.51\textwidth\epsfysize=0.23\textwidth
\epsfbox{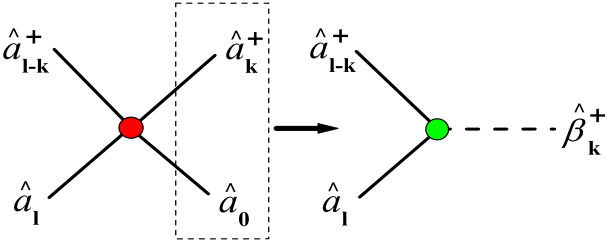}}
\caption{Two particle scattering process involving the condensate, and the
same process in terms of canonical ensemble quasiparticles.}
\label{apqe}
\end{figure}

\bigskip Consider next the conventional atom-atom interaction Hamiltonian as
sketched in Fig. \ref{apqe}. Thus we may write 
\begin{equation}
V_{1}=\sum_{\mathbf{k},\mathbf{l}}U_{\mathbf{k}\mathbf{l}}\hat{a}_{\mathbf{%
l-k}}^{\dagger }\hat{a}_{\mathbf{k}}^{\dagger }\hat{a}_{0}\hat{a}_{\mathbf{l}%
}=\sum_{\mathbf{k},\mathbf{l}}\tilde{U}_{\mathbf{k}\mathbf{l}}\hat{a}_{%
\mathbf{l-k}}^{\dagger }\hat{\beta}_{\mathbf{k}}^{\dagger }\hat{a}_{\mathbf{l%
}}  \label{V1}
\end{equation}%
where we have introduced the canonical-ensemble quasiparticles $\hat{a}_{%
\mathbf{k}}^{\dagger }{\hat{a}}_{0}\approx \sqrt{\bar{n}_{0}}\hat{a}_{%
\mathbf{k}}^{\dagger }(1+\hat{n}_{0})^{-1/2}{\hat{a}}_{0}=\sqrt{\bar{n}_{0}}%
\hat{\beta}_{\mathbf{k}}^{\dagger }$, and the notation $\tilde{U}_{\mathbf{k}%
\mathbf{l}}=\sqrt{n_{0}}U_{\mathbf{k}\mathbf{l}}$ with $U_{\mathbf{k}\mathbf{%
l}}$ being the scattering matrix element. Finally we take $\mathbf{l}=%
\mathbf{k}$ (leading term) and thus write the interaction Hamiltonian 
\begin{equation}
V=\sum_{\mathbf{k}}\tilde{U}\hat{a}_{0}^{\dagger }\hat{\beta}_{\mathbf{k}%
}^{\dagger }\hat{a}_{\mathbf{k}}+\text{adj.}  \label{V_quasi}
\end{equation}

It is important to note that this Hamiltonian involves two different kinds
of operators. The operators $\hat{a}_{0}^{\dagger }$ and $\hat{a}_{\mathbf{k}%
}$ are single particle operators which create and annihilate particles. The $%
\hat{\beta}_{\mathbf{k}}^{\dagger }$ operator, by construction conserves
particle number as befits a phonon like quasiparticle. 

\subsection{The Master Equation}

The present master equation for the condensate is now derived by focusing on
the truncated scattering Hamiltonian Eq. (\ref{V_quasi}), which adds
(removes) atoms from the condensate with the annihilation (creation) of an
excited atom and the emission (absorption) of a quasiparticle. As in CNB1
and CNB2 we treat the phonon like quasiparticle states as reservoir states
for the condensate.

The dynamical evolution of the condensate thus involves three components:
the ground state (condensate) atoms, excited (noncondensate) and the phonon
like canonical ensemble excitations, $\hat{\beta}_{\mathbf{k}}$ and $\hat{%
\beta}_{\mathbf{k}}^{\dagger }$. We proceed to treat the $\hat{\beta}_{%
\mathbf{k}}$ quasiparticles as a thermal reservoir which we trace over to
obtain the density matrix equation for the ground state, as in CNB1,2. Some
details of the master equation analysis are given in Appendix A. As will be
discussed further elsewhere, the equation of motion for the probability to
find $n_{0}$ particles in the condensate $\rho _{n_{0},n_{0}}=P_{n_{0}}$ is
now given by%
\begin{equation*}
\frac{1}{\kappa }\dot{P}%
_{n_{0}}=-K_{n_{0}}(n_{0}+1)P_{n_{0}}+K_{n_{0}-1}n_{0}P_{n_{0}-1}-
\end{equation*}%
\begin{equation}
H_{n_{0}}n_{0}P_{n_{0}}+H_{n_{0}+1}(n_{0}+1)P_{n_{0}+1}\text{.}
\label{master Pn0}
\end{equation}%
As in CNB1,2, we have divided the physical process into two kinds of terms:
the $K_{n_{0}}$ and $K_{n_{0}-1}$ terms describing the cooling of the gas
which increases the condensate number and heating terms $H_{n_{0}}$ and $%
H_{n_{0}+1}$ which decrease it. The constant $\kappa $ is an uninteresting
overall rate factor. The cooling and heating coefficients are given by 
\begin{eqnarray}
K_{n_{0}} &=&\sum_{\mathbf{k}\neq 0}\left\langle n_{\mathbf{k}}\right\rangle
_{n_{0}}\left( 1+\left\langle \hat{\beta}_{\mathbf{k}}^{\dagger }\hat{\beta}%
_{\mathbf{k}}\right\rangle \right) ,  \label{cofKn0} \\
H_{n_{0}} &=&\sum_{\mathbf{k}\neq 0}\left\langle \hat{\beta}_{\mathbf{k}%
}^{\dagger }\hat{\beta}_{\mathbf{k}}\right\rangle \left( 1+\left\langle n_{%
\mathbf{k}}\right\rangle _{n_{0}}\right) .  \label{cofHn0}
\end{eqnarray}%
Equation (\ref{master Pn0}) was obtained previously in a similar form for
the ideal gas; the essential difference being that the average $\left\langle 
\hat{\beta}_{\mathbf{k}}^{\dagger }\hat{\beta}_{\mathbf{k}}\right\rangle $
is replaced by an average number of thermal phonons in our earlier work. The
physics behind (\ref{master Pn0}) is best understood by going to a
\textquotedblleft low temperature limit\textquotedblright\ such that $%
\left\langle \hat{\beta}_{\mathbf{k}}^{\dagger }\hat{\beta}_{\mathbf{k}%
}\right\rangle \ll 1$ and we may write Eq.~(\ref{cofKn0}) as 
\begin{subequations}
\begin{equation}
K_{n_{0}}={\sum_{\mathbf{k\neq 0}}}\left\langle n_{\mathbf{k}}\right\rangle
_{n_{0}}=N-n_{0},  \label{cofKn0simple}
\end{equation}%
and $\left\langle n_{\mathbf{k}}\right\rangle _{n_{0}}\ll 1$ so that we may
write Eq.~(\ref{cofHn0}) as 
\begin{equation}
H_{n_{0}}={\sum_{\mathbf{k\neq 0}}}\left\langle \hat{\beta}_{\mathbf{k}%
}^{\dagger }\hat{\beta}_{\mathbf{k}}\right\rangle \equiv \mathcal{H}_{I}.
\label{cofHn0simple}
\end{equation}

We proceed by writing the $\hat{\beta}_{\mathbf{k}}$ and $\hat{\beta}_{%
\mathbf{k}}^{\dagger }$ operators in terms of the Bogoliubov quasiparticle
operators 
\end{subequations}
\begin{eqnarray}
\hat{\beta}_{\mathbf{k}} &=&u_{\mathbf{k}}\mathfrak{\hat{b}}_{\mathbf{k}}+v_{%
\mathbf{k}}\mathfrak{\hat{b}}_{-\mathbf{k}}^{\dagger } \\
\hat{\beta}_{-\mathbf{k}}^{\dagger } &=&u_{\mathbf{k}}\mathfrak{\hat{b}}_{-%
\mathbf{k}}^{\dagger }+v_{\mathbf{k}}\mathfrak{\hat{b}}_{\mathbf{k}}
\end{eqnarray}%
where, as usual, $[\mathfrak{\hat{b}}_{\mathbf{k}},\mathfrak{\hat{b}}_{%
\mathbf{k}^{\prime }}^{\dagger }]=\delta _{\mathbf{k},\mathbf{k}^{\prime }}$%
, hence 
\begin{eqnarray}
\left\langle \hat{\beta}_{\mathbf{k}}^{\dagger }\hat{\beta}_{\mathbf{k}%
}\right\rangle &=&u_{\mathbf{k}}^{2}\left\langle \mathfrak{\hat{b}}_{\mathbf{%
k}}^{\dagger }\mathfrak{\hat{b}}_{\mathbf{k}}\right\rangle +v_{\mathbf{k}%
}^{2}\left\langle \mathfrak{\hat{b}}_{-\mathbf{k}}\mathfrak{\hat{b}}_{-%
\mathbf{k}}^{\dagger }\right\rangle  \notag \\
&=&\frac{u_{\mathbf{k}}^{2}+v_{\mathbf{k}}^{2}}{e^{\beta \epsilon _{\mathbf{k%
}}}-1}+v_{\mathbf{k}}^{2}\equiv \mathfrak{N}_{\mathbf{k}}\text{.}  \label{uv}
\end{eqnarray}%
In the preceding we have taken the quasi-particle to be in thermal
equilibrium such that 
\begin{equation}
\left\langle \mathfrak{\hat{b}}_{\mathbf{k}}^{\dagger }\mathfrak{\hat{b}}_{%
\mathbf{k}}\right\rangle =\left\langle \mathfrak{\hat{b}}_{-\mathbf{k}%
}^{\dagger }\mathfrak{\hat{b}}_{-\mathbf{k}}\right\rangle =\frac{1}{e^{\beta
\epsilon _{\mathbf{k}}}-1}\text{.}
\end{equation}%
Finally, we note that in the useful notation of CNB3 the $\mathcal{H}_{I}$
coefficient found from Eqs. (\ref{cofHn0simple}) and (\ref{uv}) reads 
\begin{equation}
\mathcal{H}_{I}=\frac{1}{2}{\sum_{\mathbf{k\neq 0}}}\left( \frac{1}{z(A_{%
\mathbf{k}})-1}+\frac{1}{z(-A_{\mathbf{k}})-1}\right) ,
\end{equation}%
where 
\begin{equation}
z(A_{\mathbf{k}})=\frac{A_{\mathbf{k}}-e^{\varepsilon _{\mathbf{k}}/T}}{A_{%
\mathbf{k}}e^{\varepsilon _{\mathbf{k}}/T}-1}
\end{equation}%
and $A_{\mathbf{k}}$ is given by Eq. (\ref{Ak}).

In the low temperature limit the cooling and heating coefficients are given
by Eqs. (\ref{cofKn0simple}) and (\ref{cofHn0simple}). However, as was shown
in CNB2, the cross coefficient 
\begin{equation}
{\sum_{\mathbf{k\neq 0}}}\left\langle n_{k}\right\rangle
_{n_{0}}\left\langle \hat{\beta}_{\mathbf{k}}^{\dagger }\hat{\beta}_{\mathbf{%
k}}\right\rangle
\end{equation}%
as it appears in Eqs. (\ref{cofKn0}) and (\ref{cofHn0}) is necessary in
order to obtain an accurate description of the condensate. As was shown in
CNB2 and as sketched in the appendix B, for the ideal gas we have 
\begin{equation}
{\sum_{\mathbf{k\neq 0}}}\left\langle n_{k}\right\rangle
_{n_{0}}\left\langle \hat{b}_{\mathbf{k}}^{\dagger }\hat{b}_{\mathbf{k}%
}\right\rangle =(N-n_{0})\eta
\end{equation}%
where the phonon average number $\left\langle \hat{b}_{\mathbf{k}}^{\dagger }%
\hat{b}_{\mathbf{k}}\right\rangle =(e^{\beta \epsilon _{\mathbf{k}%
}}-1)^{-1}\equiv \eta _{\mathbf{k}}$ and the CNB2 cross coupling coefficient 
$\eta $ is defined by 
\begin{equation}
\eta ={\sum_{\mathbf{k\neq 0}}}\eta _{\mathbf{k}}^{2}/\mathcal{H}
\end{equation}%
where 
\begin{equation}
\mathcal{H}={\sum_{\mathbf{k\neq 0}}}\eta _{\mathbf{k}}\text{.}
\end{equation}

For the interacting gas the heating coefficient $\mathcal{H}_{I}$ and the
cross coupling coefficient can now be written as 
\begin{equation}
\mathcal{H}_{I}={\sum_{\mathbf{k\neq 0}}}\left( \frac{u_{\mathbf{k}}^{2}+v_{%
\mathbf{k}}^{2}}{e^{\beta \varepsilon _{\mathbf{k}}}-1}+v_{\mathbf{k}%
}^{2}\right) \equiv {\sum_{\mathbf{k\neq 0}}}\mathfrak{N}_{\mathbf{k}}
\label{I}
\end{equation}%
and 
\begin{equation}
{\sum_{\mathbf{k\neq 0}}}\left\langle n_{k}\right\rangle
_{n_{0}}\left\langle \hat{\beta}_{\mathbf{k}}^{\dagger }\hat{\beta}_{\mathbf{%
k}}\right\rangle ={\sum_{\mathbf{k\neq 0}}}\left\langle n_{k}\right\rangle
_{n_{0}}\mathfrak{N}_{k}=(N-n_{0})\eta _{I},  \label{N screw}
\end{equation}%
where 
\begin{equation}
\eta _{I}={\sum_{\mathbf{k\neq 0}}}\mathfrak{N}_{\mathbf{k}}^{2}/\mathcal{H}%
_{I}\text{.}  \label{etau}
\end{equation}%
Written in terms of $\mathcal{H}_{I}$ and $\eta _{I}$ the master equation
for the interacting Bose gas is finally%
\begin{equation*}
\frac{1}{\kappa }\dot{P}_{n_{0}}=-(N-n_{0})(1+\eta _{I})(n_{0}+1)P_{n_{0}}+
\end{equation*}%
\begin{equation*}
\lbrack N-(n_{0}-1)](1+\eta _{I})n_{0}P_{n_{0}-1}-[\mathcal{H}%
_{I}+(N-n_{0})\eta _{I}]n_{0}P_{n_{0}}+
\end{equation*}%
\begin{equation}
\lbrack \mathcal{H}_{I}+(N-(n_{0}+1))\eta _{I}](n_{0}+1)P_{n_{0}+1}\text{.}
\label{masteq interact}
\end{equation}

\section{Condensate Statistics via Steady State Solution to Master Equation}

\label{Sec3}

The BEC atom statistical distribution is now obtained by solving the master
equation (\ref{masteq interact}). The resulting distribution is given by
essentially the same expression as was obtained for the ideal Bose gas in
CNB2, namely%
\begin{equation}
P_{n_{0}}=\frac{1}{\mathcal{Z}_{N}}\frac{(N-n_{0}+\mathcal{H}_{I}/\eta
_{I}-1)!}{(\mathcal{H}_{I}/\eta _{I}-1)!(N-n_{0})!}\left( \frac{\eta _{I}}{%
1+\eta _{I}}\right) ^{N-n_{0}},  \label{Pn0 int}
\end{equation}%
\begin{equation}
\mathcal{Z}_{N}=\sum_{n_{0}=0}^{N}\binom{N-n_{0}+\mathcal{H}_{I}/\eta _{I}-1%
}{N-n_{0}}\left( \frac{\eta _{I}}{1+\eta _{I}}\right) ^{N-n_{0}},
\label{ZN int}
\end{equation}%
where $\mathcal{H}_{I}$ and $\eta _{I}$ for the interacting gas are given in
Table 1.

In the low temperature limit we set $\eta _{I}$ to zero and obtain the
distribution given by Eq. (\ref{rho}). The average condensate particle
number from Eq. (\ref{rho}) is 
\begin{equation}
\left\langle n_{0}\right\rangle =N-\mathcal{H}_{I}+\mathcal{H}_{I}^{N+1}/%
\mathcal{Z}_{N}N!  \label{n0 int extra}
\end{equation}%
This is in good agreement with the average value obtained in CNB3 namely $%
\left\langle n_{0}\right\rangle =N-\mathcal{H}_{I}$ valid for $T$ not in the
vicinity of $T_{c}$. The extra term $\mathcal{H}_{I}^{N+1}/\mathcal{Z}_{N}N!$
as given in Eq. (\ref{n0 int extra}) removes the unphysical cusp at $T=T_{c}$
and extends the treatment of CNB3 through the critical temperature. Equation
(\ref{rho}) and the associated $\left\langle n_{0}\right\rangle $ is
essentially the extension of the results of CNB1 to include atom-atom
interaction.

In general the $\eta_I $ parameter is important and the quasithermal $%
\left\langle n_{0}\right\rangle $ obtained from Eq. (\ref{Pn0 int}) is 
\begin{equation*}
\left\langle n_{0}\right\rangle=N-\mathcal{H}_I+P_0(\eta_I N+\mathcal{H}_I),
\end{equation*}
it is plotted in Fig. \ref{n0pqe}.

The squared variance 
\begin{equation}
\Delta n_{0}^{2}=\langle n_{0}^{2}\rangle -\langle n_{0}\rangle ^{2}
\end{equation}%
can be calculated analytically from Eq. (\ref{Pn0 int}) to find%
\begin{equation*}
\Delta n_{0}^{2}=(1+\eta _{I})\mathcal{H}_{I}-P_{0}(\eta _{I}N+\mathcal{H}%
_{I})(N-\mathcal{H}_{I}+1+\eta _{I})-
\end{equation*}%
\begin{equation}
P_{0}^{2}(\eta _{I}N+\mathcal{H}_{I})^{2},  \label{Dn0analytic}
\end{equation}%
where 
\begin{equation}
P_{0}=\frac{1}{\mathcal{Z}_{N}}\frac{(N+\mathcal{H}_{I}/\eta _{I}-1)!}{N!(%
\mathcal{H}_{I}/\eta _{I}-1)!}\left( \frac{\eta _{I}}{1+\eta _{I}}\right)
^{N}
\end{equation}%
is the probability that there are no atoms in the condensate.

If the temperature is not too close to the critical temperature, only the
first term in Eq. (\ref{Dn0analytic}) remains, resulting in 
\begin{equation}
\Delta n_{0}^{2}\simeq (1+\eta _{I})\mathcal{H}_{I}\equiv \sum\limits_{%
\mathbf{k}}(\mathfrak{N}{}_{\mathbf{k}}^{2}+\mathfrak{N}{}_{\mathbf{k}}).
\end{equation}

We plot the variance in Fig. \ref{v2pqe} for the case of 200 atoms in a box
and the values of gas parameter $an^{1/3}=0$, $0.1$. Here $a=MU/4\pi \hbar
^{2}$ is the s-wave scattering length, $M$ is the atom mass, $U\simeq U_{%
\mathbf{k}\mathbf{l}}$ is the scattering matrix element in Eq. (\ref{V1})
and $n$ is the particle density in the box. Squares show the result obtained
from CNB3, that is%
\begin{equation*}
\langle (n_{0}-\bar{n}_{0})^{2}\rangle =\frac{1}{2}\sum_{\mathbf{k}\neq 0}%
\left[ \frac{1}{(z(A_{\mathbf{k}})-1)^{2}}+\frac{1}{(z(-A_{\mathbf{k}%
})-1)^{2}}\right. +
\end{equation*}%
\begin{equation}
\left. \frac{1}{z(A_{\mathbf{k}})-1}+\frac{1}{z(-A_{\mathbf{k}})-1}\right] 
\text{,}  \label{II100}
\end{equation}%
which is in agreement with Ref. \cite{GPS}.

\begin{table}[tbp]
\begin{tabular}{|c|c|c|c|}
\hline
& $%
\begin{array}{c}
\text{low temperature} \\ 
\text{approximation}%
\end{array}
$ & $%
\begin{array}{c}
\text{quasithermal} \\ 
\text{approximation}%
\end{array}
$ & $%
\begin{array}{c}
\text{quasithermal} \\ 
\text{approximation} \\ 
\text{improved by CNB3}%
\end{array}
$ \\ \hline
$\mathcal{H}_I$ & {\large $\sum_{k\neq 0}\mathfrak{N} {}_{k}$ } & {\large $%
\sum_{k\neq 0}\mathfrak{N} {}_{k}$ } & {\large $\Delta +\sum_{k\neq 0}%
\mathfrak{N} {}_{k}$ } \\ \hline
{\large $\eta_I $ } & {\large 0 } & {\large $\frac{\sum_{k\neq 0}\mathfrak{N}
^{2}{}_{k}}{\sum_{k^{\prime }\neq 0}\mathfrak{N} {}_{k}}$ } & {\large $\frac{%
\sum_{k\neq 0}\mathfrak{N} ^{2}{}_{k}}{\sum_{k^{\prime }\neq 0}\mathfrak{N}
{}_{k}}$ } \\ \hline
\end{tabular}%
\caption{Expressions for $\mathcal{H}_I$ and $\protect\eta_I $ in three
limits of operation corresponding to the three previous treatments extended
to the present model. The average particle number in the excited level is
given by $\mathfrak{N} _{\mathbf{k}}{=}\frac{u_{\mathbf{k}}^{2}+v_{\mathbf{k}%
}^{2}}{e^{\protect\beta \protect\varepsilon _{\mathbf{k}}}-1}+v_{\mathbf{k}%
}^{2}$.}
\end{table}

Table 1 shows $\mathcal{H}_{I}$ and $\eta _{I}$ in different approximations.
The interacting Bose gas values of $\mathcal{H}_{I}$ and $\eta _{I}$ are
given in column one in the CNB1 low temperature limit in which $\eta _{I}$ =
0. The second column gives the heating $\mathcal{H}_{I}$ and cross coupling $%
\eta _{I}$ coefficients according to the Eqs. \ (\ref{I}) and (\ref{etau}).
However, we have specific information from CNB3 such that we know $\Delta
n_{0}$ etc. exactly at, say, $T=0$. We use that information to improve the
present treatment of $\Delta n_{0}$ by shifting $\mathcal{H}_{I}$ by a
constant value $\Delta $ so that we match $\Delta n_{0}$ of CNB3 at $T=0$.

\begin{figure}[h]
\bigskip 
\centerline{\epsfxsize=0.55\textwidth\epsfysize=0.45\textwidth
\epsfbox{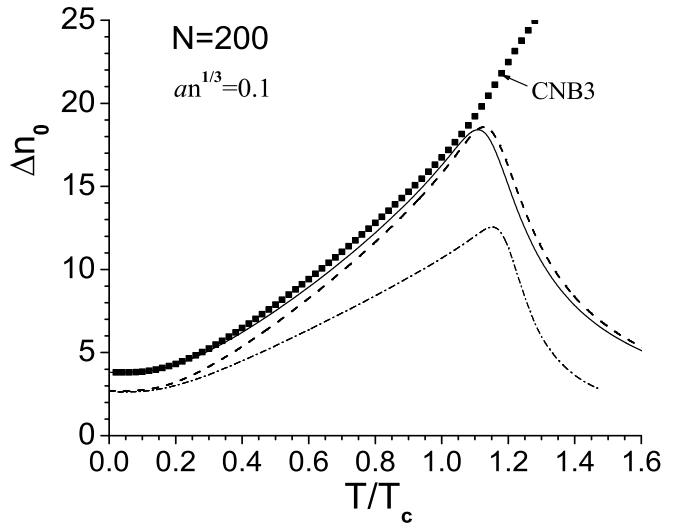}}
\caption{ Variance $\Delta n_{0}=\protect\sqrt{\langle n_{0}^{2}\rangle
-\langle n_{0}\rangle ^{2}}$ of the condensate particle number as a function
of temperature for $N=200$ particles in a box calculated via the solution of
the condensate master equation for $an^{1/3}=0.1$ in different
approximations. Dash-dot line is obtained in the low temperature limit.
Dashed line is the quasithermal approximation. Squares show the result of
Ref. \protect\cite{GPS} and CNB3. Solid line is calculated in the
quasithermal approximation by a shift of $\mathcal{H}_I$ to match the CNB3
value at $T=0$. }
\label{v2pqe1}
\end{figure}

Figure \ref{v2pqe1} compares the variance $\Delta n_{0}$ of the condensate
particle number calculated for $an^{1/3}$ = 0.1 in three different
approximations. The dash-dot line is obtained in the low temperature limit
(Table 1 column one). The dashed line is the quasithermal approximation
(Table 1 column two). The solid curve was obtained from $\mathcal{H}_I$ and $%
\eta_I $ of Table 1 column three.

\section{Discussion and Summary}

\subsection{Mean number of particles in the condensate}

The calculation of $\langle n_{0}\rangle $ for the interacting
``Bose-Bogoliubov'' gas is in good agreement with and extends the results of
CNB3. For example, the Uhlenbeck cusp dilemma \cite{Uhlenbeck} is resolved
for the interacting gas just as it was for the ideal gas. The master
equation approach gives an excellent treatment of the temperature dependence
average number of atoms in the condensate. The agreement with the rigorous
results of CNB3 is gratifying. Note in particular the ground state depletion
at $T=0$, due to atom-atom interactions, is handled very well by the present
approach. Likewise the increase in the number of atoms in the ground state,
for temperature $0.3\lesssim T/T_{c}\lesssim 1.0$, is modelled very well by
the master equation approach. For temperatures well above $T_{c}$ ($%
T/T_{c}\gtrsim 1.4$) the ideal gas treatment is valid and agrees with our
result. Furthermore, and most significantly from the vantage point of this
paper, the master equation approach works well for the temperature region
near $T_{c}$ where the other treatments fail.

As one can see from Fig. \ref{n0pqe}, repulsive interaction between Bose
particles stimulates BEC and yields an increase in $\bar{n}_{0}$ at
intermediate temperature as compared to the ideal gas \cite{Leggett}. The
reason is energetic: bosons in different but mutually overlapping states
interact stronger than when they are in the same state. For example, when
two particles are in the same state $\Psi (\mathbf{r}_{1},\mathbf{r}%
_{2})=\chi (\mathbf{r}_{1})\chi (\mathbf{r}_{2})$ and interact via a
potential $V=g\delta (\mathbf{r}_{1}-\mathbf{r}_{2})$, the interaction
energy is 
\begin{equation*}
E_{\text{int}}=\int d\mathbf{r}_{1}\int d\mathbf{r}_{2}\Psi ^{\ast }(\mathbf{%
r}_{1},\mathbf{r}_{2})V\Psi (\mathbf{r}_{1},\mathbf{r}_{2})=
\end{equation*}
\begin{equation*}
g\int d\mathbf{r}_{1}|\chi (\mathbf{r}_{1})|^{4}.
\end{equation*}
However when two particles are in different states such that $\Psi (\mathbf{r%
}_{1},\mathbf{r}_{2})=[\varphi (\mathbf{r}_{1})\chi (\mathbf{r}_{2})+\varphi
(\mathbf{r}_{2})\chi (\mathbf{r}_{1})]/\sqrt{2}$, the interaction energy is 
\begin{equation*}
E_{\text{int}}=2g\int d\mathbf{r}_{1}|\varphi (\mathbf{r}_{1})|^{2}|\chi (%
\mathbf{r}_{1})|^{2}.
\end{equation*}
Then since $|\varphi (\mathbf{r}_{1})|^{2}\simeq |\chi (\mathbf{r}_{1})|^{2}$
we see that two bosons in the same state is the lowest energy configuration.
This favors multiple occupation of a single one-particle state. Such an
effect is sometimes called an attraction in momentum space \cite{Huang}. One
can see from Fig. \ref{n0pqe} that repulsive interaction also increases $T_c$%
.

\subsection{Fluctuations $\Delta n_{0}$}

As Einstein taught us long ago, fluctuation phenomena contains much more
physics then mean values do. He used the difference between the fluctuation
properties of waves and particles to show the particle side of the photon 
\cite{E05} and the wave side of matter \cite{E25}. Note that the latter was
well in advance of the Schr\"{o}dinger equation, and provided support for
the wave nature of matter.

\begin{figure}[h]
\bigskip 
\centerline{\epsfxsize=0.55\textwidth\epsfysize=0.45\textwidth
\epsfbox{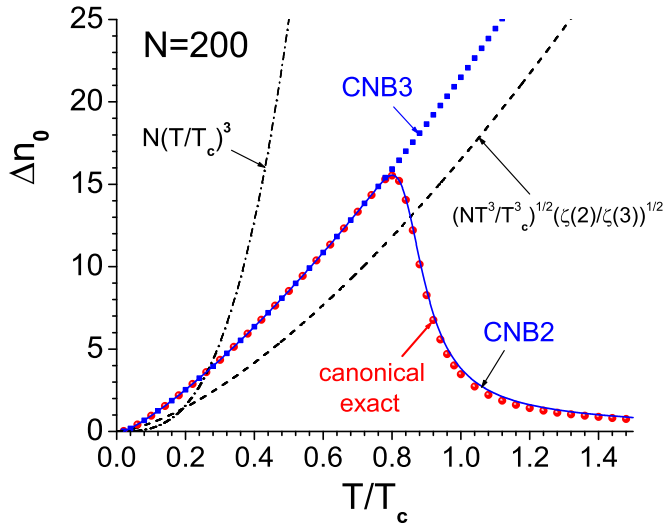}}
\caption{Variance $\Delta n_{0}=\protect\sqrt{\langle n_{0}^{2}\rangle
-\langle n_{0}\rangle ^{2}}$ of the condensate particle number as a function
of temperature for an ideal Bose gas of $N=200$ atoms in an isotropic
harmonic trap. Solid line is a solution of the condensate master equation.
Dots are the exact numerical results obtained in the canonical ensemble for
the ideal Bose gas \protect\cite{la}. Squares are result of CNB3. Dashed
line is a plot of $\Delta n_{0}=\protect\sqrt{\frac{\protect\zeta (2)N}{%
\protect\zeta (3)}\left( \frac{T}{T_{c}}\right) ^{3}}$ which is obtained in
the thermodynamic limit \protect\cite{Politzer}. Dash-dot line is a plot of $%
\Delta n_{0}=N-\langle n_{0}\rangle =N\left( \frac{T}{T_{c}}\right) ^{3}$,
which is proposed by D. ter Haar \protect\cite{Haar} in the low temperature
regime (adapted to a harmonic trap). This had the correct zero limit as $%
T\rightarrow 0$, but is not right for higher temperatures.}
\label{v2pqe2}
\end{figure}

Likewise fluctuations are central to our investigation. We find that even in
the ideal Bose gas the fluctuation physics is quite rich, see Fig. \ref%
{v2pqe2}. The atomic statistical distribution, and its first two central
moments, as plotted in Fig. \ref{n0pqe} (for $\bar{n}_{0}$) and Fig. \ref%
{v2pqe} (for $\Delta n_{0}$) make our point. The low temperature
approximation used to get Eq. (\ref{rho}) works quantitatively well for $%
\bar{n}_{0}$. However, the treatment of fluctuations in this approximation
is only qualitative (see Fig. \ref{v2pqe1}). To get a good description of
fluctuations we must extend the improved approach of CNB2, with appropriate
modifications, to the case of the interacting gas.

The extension of the ideal gas work of CNB2, as per column 2 of Table 1 to
the present problem yields the dashed curve which is better but not good for 
$T=0$. However, CNB3 provides a good description of the central moments $%
\langle (n_{0}-\bar{n}_{0})^{m}\rangle $ for low temperatures. Thus we can
improve our description of the fluctuations by using the known low
temperature behavior from CNB3 to improve the heating coefficients as in
column 3 of Table 1. The utilization of the low temperature results of CNB3
to improve on $K_{n_{0}}$ and $H_{n_{0}}$ yields an excellent description of
the central moments, as will be shown elsewhere.

\subsection{Connection with previous studies and future work}

The experiments in dilute Bose gases of rubidium, lithium and sodium have
reopened the old BEC questions, e.g., the Uhlenbeck cusp dilemma \cite%
{Uhlenbeck} and the question of condensate fluctuations. See for example the
recent review \cite{review} and references therein. Likewise the condensate
time development \cite{Mies98} is an exciting frontier.

In earlier work \cite{MOS99}, \cite{KSZZ} we found that there is a useful
connection between BEC of an ideal Bose gas \cite{Bogo87}, and the quantum
theory of the laser \cite{Hake,Holl}. We recall that the saturation
nonlinearity in the radiation matter interaction is essential for laser
coherence \cite{Glau63}.

Indeed, the coherence generating nonlinearity in the case of the ideal Bose
gas is the particle number constraint which provides the essential
nonlinearity in BEC. However, in the case of the interacting gas, the
problem of fluctuation is rather subtle and requires a more refined
approach. Although useful papers have been published dealing with various
limiting cases \cite{GPS, Dunningham}, so far there has been no treatment of
this problem valid at all temperatures. Apart from general theoretical
interest, condensate fluctuations can be measured, in principle, by means of
a scattering of series of short pulses \cite{Idzia}, see also \cite{Chuu05}.

The problem of $N$ ideal bosons in a 3D harmonic potential coupled to a
thermal reservoir turns out to be surprisingly rich. The $N$ particle
constraint is included naturally in the present formulation and introduces
an essential nonlinearity yielding accurate results as shown in Fig. \ref%
{n0pqe} and \ref{v2pqe}.

Given the utility of the quantum theory of the laser-master equation
approach in the ideal Bose gas problem it is natural to look at the
interacting gas from a quantum optical perspective, and that was the vantage
of Ref. \cite{KKS}. Other fascinating studies in this regard include the
paper by Lewenstein and You on quantum phase diffusion \cite{LewensteinYou},
Graham \cite{Graham} who applied quantum noise analysis to the BEC linewidth
question and Wiseman and Thomson on reducing the linewidth of an atom laser
by feedback \cite{WisemanThomsen}. The time dependent master equation
analysis of Gardiner and Zoller is presented in a heroic series of strong
papers \cite{GZ}. The paper of Parkins and Walls \cite{ParkinsWalls} on the
BEC is another important work in this field. Reference to many other useful
and insightful papers is given in our recent review on fluctuations in the
BEC \cite{review}.

One basic difference between the classic papers mentioned above is treatment
of the BEC of a mesoscopic gas in the critical region near $T_{c}$. The
present master equation approach is, to our knowledge, unique in that it
treats BEC analytically at all temperatures.

In the next paper (CNB5) we shall follow the lead of Section \ref{Sec3}, see
Table I and Fig. \ref{v2pqe1}, wherein we use the statistical mechanical
results of CNB3 to find a good (semi-phenomenological) description of all
moments at arbitrary temperatures. Then in CNB6 we shall present a detailed
analysis including the off-diagonal elements (e.g. $\hat{a}_{0}^{\dagger }%
\hat{a}_{0}^{\dagger }\hat{\rho}_{0}$). Finally we will return to the study
of various forms of the dynamics (Hamiltonian) in CNB7, and show that
certain formulations do not involve off-diagonal elements and give a good
account of the physics via a simple treatment.

\section{Acknowledgments}

We wish to thank Vitaly Kocharovsky for useful discussions. We gratefully
acknowledge the support of the Office of Naval Research (Award No.
N00014-03-1-0385) and the Robert A. Welch Foundation (Grant No. A-1261).

\appendix

\section{The Master Equation}

\bigskip In CNB1,2 the equation of motion for the total \textquotedblleft
gas+reservoir\textquotedblright\ system density matrix in the interaction
representation was given by 
\begin{equation}
\dot{\rho}_{\text{total}}(t)=-i[V(t),\rho _{\text{total}}(t)]/\hbar \text{.}
\end{equation}%
Integrating for $\rho _{\text{total}}$, inserting it back into the equation
of motion for $\dot{\rho}(t)_{\text{total}}$, and tracing over the
reservoir, we obtain a useful equation of motion for the density matrix of
the Bose-gas subsystem 
\begin{equation}
\dot{\rho}(t)=-\frac{1}{\hbar ^{2}}\int_{0}^{t}dt^{\prime }\mathrm{Tr}_{%
\mathrm{res}}[V(t),[V(t^{\prime }),\rho _{\text{total}}(t^{\prime })]],
\end{equation}%
where $\mathrm{{Tr}_{res}}$ denotes the trace over the $\hat{\beta}_{\mathbf{%
k}}$ degrees of freedom.

We proceeded by assuming that the phonon reservoir (in the spirit of the
experiments of Reppy and coworkers \cite{rep}) remains unchanged during the
interaction with the Bose gas. As discussed in detail in CNB1,2, the density
operator for the total system \textquotedblleft
gas+reservoir\textquotedblright\ can then be factored, i.e., $\rho _{\text{%
total}}(t^{\prime })\approx \rho (t^{\prime })\otimes \rho _{\text{res}}$,
where $\rho _{\text{res}}$ is the equilibrium density matrix of the
reservoir. If the spectrum is smooth, we are justified in making the Markov
approximation, viz.\ $\rho (t^{\prime })\rightarrow \rho (t)$.

In the present analysis we do not invoke an external (phonon) reservoir.
Instead we regard the excited states of our weakly interacting gas to be
reservoir-like. In this regard the excited states are nearly analogous to
the atoms in the quantum theory of the laser \cite{Hake}. There we calculate
the change in the laser radiation density matrix due to one atom and
multiply by the rate of such atomic interaction to obtain a coarse-grained
equation of motion for the laser density matrix.

Following the laser approach explicitly (in CNB1,2 it was only implicitly)
we first calculate $\delta \rho (\hat{a}_{0},\hat{a}_{0}^{\dagger },t)$ due
to a collision between two atoms, which adds (or removes) atoms to the
ground state and then multiply by the rate of such collisions to find our
master equation for $\rho (\hat{a}_{0},\hat{a}_{0}^{\dagger },t)$. The small
change $\delta \rho $ is given by%
\begin{equation}
\delta \rho (\hat{a}_{0},\hat{a}_{0}^{\dagger },t)=-\frac{1}{\hbar ^{2}}%
\int\limits_{t}^{t+\tau _{c}}dt^{\prime }\int\limits_{t}^{t+t^{\prime
}}dt^{\prime \prime }\mathrm{{Tr}_{exc}}[V(t^{\prime }),[V(t^{\prime \prime
}),\rho (t)]],  \label{A10}
\end{equation}%
where the collision time $\tau _{c}$ is of the order of the scattering
length divided by the thermal velocity and is some nanoseconds in duration.
The excited states, $\{n_{\mathbf{k}}\}$, are not much influenced by a
single collision and the density matrix $\rho (t)$ is taken to be $\rho
(t)=\rho _{\text{res}}(t)\otimes \rho (\hat{a}_{0},\hat{a}_{0}^{\dagger },t)$%
. It is important to note that the trap frequencies are of order a few
hundred Hertz and the integrals over $t^{\prime }$ and $t^{\prime \prime }$
can be simply replaced by $\tau _{c}^{2}/2$. The rate of collisions $r$ is
governed by the particle mean free path, the thermal velocity and the number
of (density of) excited states. The rate $\ r$ will contribute only to the
constant $\kappa $ of e.g. Eq. (\ref{master eqt}) and is not of interest
here.

Using Eq. (\ref{A10}) and the interaction Hamiltonian Eq. (\ref{V_quasi}) we
obtain the following coarse grained equation of motion for the reduced
density operator of the interacting Bose gas%
\begin{equation*}
\dot{\rho}(\hat{a}_{0},\hat{a}_{0}^{\dagger },t)=r\delta \rho =-\frac{\kappa 
}{2}\sum\limits_{\mathbf{k\neq 0}}\mathrm{Tr}\{\hat{\beta}_{\mathbf{k}}\hat{%
\beta}_{\mathbf{k}}^{\dagger }\rho _{res}\}[\hat{a}_{\mathbf{k}}^{\dagger }%
\hat{a}_{\mathbf{k}}\hat{a}_{0}\hat{a}_{0}^{\dagger }\hat{\rho}(t)-
\end{equation*}%
\begin{equation*}
\hat{a}_{\mathbf{k}}\hat{a}_{0}^{\dagger }\hat{\rho}(t)\hat{a}_{0}\hat{a}_{%
\mathbf{k}}^{\dagger }+\text{adj.}]-
\end{equation*}%
\begin{equation}
\frac{\kappa }{2}\sum\limits_{\mathbf{k\neq 0}}\mathrm{Tr}\{\hat{\beta}_{%
\mathbf{k}}^{\dagger }\hat{\beta}_{\mathbf{k}}\rho _{res}\}[\hat{a}_{\mathbf{%
k}}\hat{a}_{\mathbf{k}}^{\dagger }\hat{a}_{0}^{\dagger }\hat{a}_{0}\hat{\rho}%
(t)-\hat{a}_{\mathbf{k}}^{\dagger }\hat{a}_{0}\hat{\rho}(t)\hat{a}%
_{0}^{\dagger }\hat{a}_{\mathbf{k}}+\text{adj.}].  \label{master eqt}
\end{equation}%
The constant $\kappa $ is an overall rate which plays no role in the present
problem.

In this paper, we are most interested in the atomic statistics of the
condensate and hence define the joint probability for having $n_{0}$ atoms
in the condensate and $n_{1},n_{2},$.., $n_{\mathbf{k}}$,..$=\{n_{\mathbf{k}%
}\}$ atoms in the various excited states as: 
\begin{equation}
P(\{n_{\mathbf{k}}\},n_{0})=\langle \{n_{\mathbf{k}}\},n_{0}|\rho |\{n_{%
\mathbf{k}}\},n_{0}\rangle  \label{P noncondensate}
\end{equation}

In view of the definition (\ref{P noncondensate}), Eq.~(\ref{master eqt})
implies%
\begin{equation*}
\dot{P}(\{n_{\mathbf{k}}\};n_{0})=-\frac{\kappa }{2}\sum\limits_{\mathbf{%
k\neq 0}}(\langle \hat{\beta}_{\mathbf{k}}^{\dagger }\hat{\beta}_{\mathbf{k}%
}\rangle +1)[n_{\mathbf{k}}(n_{0}+1)P(\{n_{\mathbf{k}}\};n_{0})-
\end{equation*}%
\begin{equation*}
n_{\mathbf{k}}n_{0}P(\{n_{\mathbf{k}}\};n_{0}-1)+\text{adj.}]-
\end{equation*}%
\begin{equation*}
\frac{\kappa }{2}\sum\limits_{\mathbf{k\neq 0}}\langle \hat{\beta}_{\mathbf{k%
}}^{\dagger }\hat{\beta}_{\mathbf{k}}\rangle \lbrack (n_{\mathbf{k}%
}+1)n_{0}P(\{n_{\mathbf{k}}\};n_{0})-
\end{equation*}%
\begin{equation}
(n_{\mathbf{k}}+1)(n_{0}+1)P(\{n_{\mathbf{k}}\};n_{0}+1)+\text{adj.}]
\end{equation}

The condensate statistical distribution $P(n_{0})$ is then related to the
joint probability distribution $P(\{n_{\mathbf{k}}\};n_{0})$ and the
conditional probability $P(\{n_{\mathbf{k}}\}|n_{0})$ via Bayes rule 
\begin{equation}
P(\{n_{\mathbf{k}}\};n_{0})=P(\{n_{\mathbf{k}}\}|n_{0})P(n_{0})\text{.}
\label{Pnkn0}
\end{equation}

We obtain $P(n_{0})$ from $P(\{n_{\mathbf{k}}\};n_{0})$ directly by summing
over all excited states $\{n_{\mathbf{k}}\}$%
\begin{equation}
P(n_{0})=\sum\limits_{\{n_{\mathbf{k}}\}}P(\{n_{\mathbf{k}}\};n_{0}),
\label{Pn0}
\end{equation}%
so that we have%
\begin{equation*}
\dot{P}_{n_{0}}=-\kappa \sum\limits_{\mathbf{k\neq 0}}(\langle \hat{\beta}_{%
\mathbf{k}}^{\dagger }\hat{\beta}_{\mathbf{k}}\rangle +1)[\langle n_{\mathbf{%
k}}\rangle _{n_{0}}(n_{0}+1)P_{n_{0}}-
\end{equation*}%
\begin{equation*}
\langle n_{\mathbf{k}}\rangle _{n_{0}-1}n_{0}P_{n_{0}-1}]-
\end{equation*}%
\begin{equation}
\kappa \sum\limits_{\mathbf{k\neq 0}}\langle \hat{\beta}_{\mathbf{k}%
}^{\dagger }\hat{\beta}_{\mathbf{k}}\rangle \lbrack (\langle n_{\mathbf{k}%
}\rangle _{n_{0}}+1)n_{0}P_{n_{0}}-(\langle n_{\mathbf{k}}\rangle
_{n_{0}+1}+1)(n_{0}+1)P_{n_{0}+1}]  \label{eqPn0}
\end{equation}%
where we have introduced the convenient conditional average notation%
\begin{equation}
\langle n_{\mathbf{k}}\rangle _{n_{0}}=\sum\limits_{n_{\mathbf{k}}}n_{%
\mathbf{k}}P(\{n_{\mathbf{k}}\}|n_{0})\text{.}
\end{equation}

The master equation obtained here is diagonal. More general master equation
can be off-diagonal, as we discuss in Appendix C.

\section{The Cross Coupling Coefficients}

In this appendix, we present a brief reminder of how the heating and cooling
coefficients $H_{n_{0}}$ and $K_{n_{0}}$ were handled in CNB2 for an ideal
gas. There, we encountered the heating and cooling coefficients 
\begin{equation}
K_{n_{0}}=\sum_{\mathbf{k\neq 0}}\left\langle n_{\mathbf{k}}\right\rangle
_{n_{0}}\left( 1+\left\langle \hat{b}_{\mathbf{k}}^{\dagger }\hat{b}_{%
\mathbf{k}}\right\rangle \right) ,  \label{Kn0 2}
\end{equation}%
and 
\begin{equation}
H_{n_{0}}=\sum_{\mathbf{k\neq 0}}\left\langle \hat{b}_{\mathbf{k}}^{\dagger }%
\hat{b}_{\mathbf{k}}\right\rangle \left( 1+\left\langle n_{\mathbf{k}%
}\right\rangle _{n_{0}}\right) .  \label{Hn0 2}
\end{equation}

Defining 
\begin{equation}
{\eta }_{k}=\langle \hat{b}_{\mathbf{k}}^{\dagger }\hat{b}_{\mathbf{k}%
}\rangle =\frac{1}{e^{\beta \hbar \omega _{k}}-1},  \label{eta}
\end{equation}%
where $\beta =1/k_{B}T$. Eqs.~(\ref{Kn0 2}) and (\ref{Hn0 2}) now read 
\begin{equation}
K_{n_{0}}=\sum_{\mathbf{k\neq 0}}\langle n_{k}\rangle _{n_{0}}(1+\eta _{k}),
\label{K4}
\end{equation}%
and 
\begin{equation}
H_{n_{0}}=\sum_{\mathbf{k\neq 0}}\eta _{k}(1+\langle n_{k}\rangle _{n_{0}}).
\label{H4}
\end{equation}%
We then approximate the conditional thermal average as 
\begin{equation}
\langle n_{k}\rangle _{n_{0}}=(N-n_{0})\frac{{\bar{n}}_{k}}{\sum_{\mathbf{%
k\neq 0}}{\bar{n}}_{k}},  \label{nkave}
\end{equation}%
where the usual atomic thermal average is given by 
\begin{equation}
{\bar{n}}_{k}=\frac{1}{e^{\beta \epsilon _{k}}-1}.  \label{nkbar}
\end{equation}%
Now an ensemble of atoms and phonons in an ordinary gas obeys the simple
rate equation 
\begin{equation}
\frac{d}{dt}{\bar{n}}_{k}=\gamma ({\bar{n}}_{k}+1){\eta }_{k}-\gamma {\bar{n}%
}_{k}({\eta }_{k}+1),  \label{nkbardot}
\end{equation}%
where $\gamma $ is an uninteresting rate factor. At steady state Eq. (\ref%
{nkbardot}) yields 
\begin{equation}
({\bar{n}}_{k}+1){\eta }_{k}={\bar{n}}_{k}({\eta }_{k}+1),  \label{te}
\end{equation}%
which implies that ${\eta }_{k}={\bar{n}}_{k}$ and we may write Eq. (\ref%
{nkave}) as 
\begin{equation}
\langle n_{k}\rangle _{n_{0}}=(N-n_{0})\frac{{\eta }_{k}}{\sum_{\mathbf{%
k\neq 0}}{\eta }_{k}}.  \label{nkave2}
\end{equation}%
Hence we have 
\begin{equation}
\sum_{\mathbf{k\neq 0}}\langle n_{k}\rangle _{n_{0}}{\eta }_{k}=\frac{%
(N-n_{0})}{\mathcal{H}}\sum_{\mathbf{k\neq 0}}{\eta }_{k}^{2}=(N-n_{0})\eta
\label{simp}
\end{equation}%
where we have introduced the notations 
\begin{equation}
\mathcal{H}=\sum_{\mathbf{k\neq 0}}\eta _{k}\text{ and }\eta =\frac{\sum_{%
\mathbf{k\neq 0}}{\eta }_{k}^{2}}{\mathcal{H}}.
\end{equation}%
In this way we arrive at the $K_{n_{0}}$ and $H_{n_{0}}$ coefficients of
CBN2 given by 
\begin{equation}
K_{n_{0}}=(N-n_{0})(1+\eta ),  \label{K5}
\end{equation}%
and 
\begin{equation}
H_{n_{0}}=\mathcal{H}+(N-n_{0})\eta .  \label{H5}
\end{equation}

\section{Master Equation Generalizations}

Our interaction Hamiltonian can yield an off-diagonal master equation. That
is, the main working master equation in operator form is%
\begin{equation*}
\frac{1}{\kappa }\frac{d\hat{\rho}(t)}{dt}=-\sum\limits_{\mathbf{k\neq 0}}<%
\hat{a}_{\mathbf{k}}^{\dagger }\hat{a}_{\mathbf{k}}\hat{\beta}_{\mathbf{k}}%
\hat{\beta}_{\mathbf{k}}^{\dagger }>_{n_{0}}(\hat{a}_{0}\hat{a}_{0}^{\dagger
}\hat{\rho}(t)-\hat{a}_{0}^{\dagger }\hat{\rho}(t)\hat{a}_{0})+\text{adj.}
\end{equation*}%
\begin{equation*}
-\sum\limits_{\mathbf{k\neq 0}}<\hat{\beta}_{\mathbf{k}}^{\dagger }\hat{\beta%
}_{\mathbf{k}}\hat{a}_{\mathbf{k}}\hat{a}_{\mathbf{k}}^{\dagger }>_{n_{0}}(%
\hat{a}_{0}^{\dagger }\hat{a}_{0}\hat{\rho}(t)-\hat{a}_{0}\hat{\rho}(t)\hat{a%
}_{0}^{\dagger })+\text{adj.}-
\end{equation*}%
\begin{equation*}
-\sum\limits_{\mathbf{k\neq 0}}<\hat{\beta}_{\mathbf{k}}^{\dagger }\hat{\beta%
}_{-\mathbf{k}}^{\dagger }\hat{a}_{\mathbf{k}}\hat{a}_{-\mathbf{k}}>_{n_{0}}(%
\hat{a}_{0}^{\dagger }\hat{a}_{0}^{\dagger }\hat{\rho}(t)-\hat{a}%
_{0}^{\dagger }\hat{\rho}(t)\hat{a}_{0}^{\dagger })+\text{adj.}-
\end{equation*}%
\begin{equation}
-\sum\limits_{\mathbf{k\neq 0}}<\hat{\beta}_{\mathbf{k}}\hat{\beta}_{-%
\mathbf{k}}\hat{a}_{\mathbf{k}}^{\dagger }\hat{a}_{-\mathbf{k}}^{\dagger
}>_{n_{0}}(\hat{a}_{0}\hat{a}_{0}\hat{\rho}(t)-\hat{a}_{0}\hat{\rho}(t)\hat{a%
}_{0})+\text{adj.}  \label{drho/dt}
\end{equation}%
New contribution of the form $\hat{a}_{0}^{\dagger }\hat{a}_{0}^{\dagger }%
\hat{\rho}_{0}(t)-\hat{a}_{0}^{\dagger }\hat{\rho}(t)\hat{a}_{0}^{\dagger }$
and $\hat{a}_{0}\hat{a}_{0}\hat{\rho}(t)-\hat{a}_{0}\hat{\rho}(t)\hat{a}_{0}$
yields off-diagonal terms in the density matrix equation. Taking into
account that 
\begin{equation*}
<\hat{\beta}_{\mathbf{k}}^{\dagger }\hat{\beta}_{-\mathbf{k}}^{\dagger }>=<%
\hat{a}_{\mathbf{k}}\hat{a}_{-\mathbf{k}}>=u_{k}v_{k}\left[ \frac{2}{%
e^{\beta \epsilon _{k}}-1}+1\right] 
\end{equation*}%
we obtain a master equation which couples the diagonal and off-diagonal
terms in the density matrix%
\begin{equation*}
\frac{1}{2\kappa }\dot{\rho}_{n_{0},n_{0}}=-(N-n_{0})(1+\eta
_{I})(n_{0}+1)\rho _{n_{0},n_{0}}+
\end{equation*}%
\begin{equation*}
+(N-n_{0}+1)(1+\eta _{I})n_{0}\rho _{n_{0}-1,n_{0}-1}-
\end{equation*}%
\begin{equation*}
-[\mathcal{H}_{I}+(N-n_{0})\eta _{I}]n_{0}\rho _{n_{0},n_{0}}+
\end{equation*}%
\begin{equation*}
+[\mathcal{H}_{I}+(N-n_{0}-1)\eta _{I}](n_{0}+1)\rho _{n_{0}+1,n_{0}+1}-
\end{equation*}%
\begin{equation*}
-J\left[ \sqrt{n_{0}(n_{0}-1)}\rho _{n_{0}-2,n_{0}}+\sqrt{(n_{0}+2)(n_{0}+1)}%
\rho _{n_{0},n_{0}+2}\right. -
\end{equation*}%
\begin{equation*}
\left. -2\sqrt{n_{0}(n_{0}+1)}\rho _{n_{0}-1,n_{0}+1}\right] 
\end{equation*}%
\begin{equation*}
-J\left[ \sqrt{(n_{0}+1)(n_{0}+2)}\rho _{n_{0}+2,n_{0}}+\sqrt{n_{0}(n_{0}-1)}%
\rho _{n_{0},n_{0}-2}\right. -
\end{equation*}%
\begin{equation}
\left. -2\sqrt{n_{0}(n_{0}+1)}\rho _{n_{0}+1,n_{0}-1}\right] ,
\end{equation}%
where $\mathcal{H}_{I}$ and $\eta _{I}$ are given by previous formulas (\ref%
{I}) and (\ref{etau}), and 
\begin{equation}
J=\sum\limits_{\mathbf{k\neq 0}}u_{k}^{2}v_{k}^{2}\left[ \frac{2}{e^{\beta
\epsilon _{k}}-1}+1\right] ^{2}.
\end{equation}

The more general analysis of fluctuations with off-diagonal terms will be
presented elsewhere. We have also studied a two atom master equation in
which two atoms at a time are added or removed from the condensate. This is
analogous to the two photon laser. The results of such an analysis are in
basic agreement with the present findings and will be published elsewhere.



\begin{thebibliography}{99}
\bibitem{Glau63} Pure coherent light is characterized by Poissonian photon
statistics. The quantum theory of the laser shows that partially coherent
laser light also has a sharp photon distribution (with width several times
Poissonian for a typical He-Ne laser). Laser coherence is due to the fact
that after emission the atoms become absorbers and this saturation
nonlinearity is essential for producing a sharply peaked photon distribution.
\end{thebibliography}
\end{document}